\documentstyle[aps,12pt,epsf]{revtex}                                           \newcommand{\etal}{\mbox{\it et al.}}
\tighten

\begin{document}

\preprint{~}

\title{Measurements of the Suitability of Large Rock Salt Formations\\
for Radio Detection of High Energy Neutrinos}

\author{
Peter~Gorham$^{1}$,
David~Saltzberg$^{2}$, 
Allen~Odian$^{3}$,
Dawn~Williams$^{2}$,\\
David~Besson$^{4}$,
George~Frichter$^{5}$,
\& Sami~Tantawi$^{3}$}
\address{$^{1}$Jet Propulsion Laboratory, Calif. Institute of Technology,
Pasadena, CA, 91109}
\address{$^{2}$Department of Physics and Astronomy,
University of California, Los Angeles, CA 90095}
\address{$^{3}$Stanford Linear Accelerator Center, Stanford University, 
Stanford, CA 94309}
\address{$^{4}$Dept. of Physics and Astronomy, University of Kansas, Lawrence, KS 66045}
\address{$^{5}$Department of Physics, Florida State University}

\date{\today}

\maketitle

\begin{abstract}

We have investigated the possibility that large rock salt formations 
might  be suitable as target masses for detection of neutrinos
of energies about 10~PeV and above.   In neutrino interactions at these
energies, the  secondary electromagnetic cascade produces a coherent
radio pulse well above ambient thermal noise via the Askaryan
effect.  We describe measurements of radio-frequency 
attenuation lengths  and ambient thermal noise in two salt formations.
Measurements in the Waste Isolation Pilot Plant (WIPP), located in an 
evaporite salt bed in Carlsbad, NM yielded short  attenuation lengths, 3--7~m
over 150--300~MHz.  However, measurements at United Salt's
Hockley mine, located in a salt dome near Houston, Texas yielded
attenuation lengths in excess of 250~m at similar frequencies. 
We have also analyzed early
ground-penetrating radar data at Hockley mine and have found 
additional evidence for attenuation lengths in 
excess of several hundred  meters at 440~MHz.   We conclude
that salt domes, which may individually contain several hundred
cubic kilometer water-equivalent mass, provide attractive sites
for next-generation high-energy neutrino detectors.

\end{abstract}
~\vspace*{-0.5in}\\



\section{Introduction}

The observation of several dozen single cosmic ray particles
with energies beyond the $~\sim 10^{19.5}$~eV
Greisen-Zatsepin-Kuzmin (GZK)~\cite{GZK} cutoff poses among the most
intriguing mysteries of high energy astrophysics. If our
local region of the universe is not atypical, the detection of these
particles implies a corresponding flux of 
$10^{17-19}$~eV neutrinos~\cite{GZKnu}. 
These neutrinos
are secondary particles of the interactions of the super-GZK cosmic-rays,
whose energy rapidly degrades over scales of a few tens of Mpc due
to photomeson production on the 3K microwave 
background.~\cite{Photomeson-old} Detection of such
neutrinos would
provide unique information about the production and
propagation of the particles near and above the GZK cutoff.
This information is almost certainly necessary and in some
cases sufficient to determine the nature of the primary
cosmic rays and their sources. Characterization of these
neutrinos is thus arguably of equal priority to the
measurement of the super-GZK cosmic-ray spectrum.

Given standard estimates of the fluxes of the 
(presumably isotropic) GZK neutrinos~\cite{GZKnu},
their detection will require
of order several tens of cubic kilometers of water-equivalent 
instrumented mass with an acceptance solid angle of $\sim 1$ sr, 
and at least several years of operation,
assuming no background.  A detector must also have reasonable
calorimetric capability in order to establish the neutrino spectral
signature of the GZK process. Present approaches, such
as the water Cherenkov detectors AMANDA~\cite{AMANDA}
and ANTARES~\cite{ANTARES}, and the planned
ICECUBE~\cite{ICECUBE}, are on a path which can achieve detection of these
neutrinos through muon tracking techniques and detection of high energy
cascades. 
The RICE project\cite{fri96,Bes99,Bes01},
utilizing cascade radio emission to search for such interactions, 
provides an alternative
radio-Cherenkov detection method, also in Antarctic ice, as
part of the AMANDA project. Both the optical and radio approaches will require significantly larger volumes to approach sensitivity to GZK neutrinos.
Here we describe measurements
which indicate the viability of an alternative medium to
ice, that of subterranean natural salt formations, which are abundant
throughout the world. As we will show here, salt formations exist with
dielectric properties that are competitive with polar sheet ice,
and these may thus provide a cost-effective approach to the 
challenging problem of GZK neutrino detection.

The first suggestion that naturally occurring rock salt formations
could be a viable neutrino detection medium can be traced back
to G. A. Askaryan\cite{Ask62}, who first proposed the 
coherent Cherenkov mechanism for radio pulse production by
charge asymmetry in electromagnetic cascades. The asymmetry
arises from photon and electron scattering processes on
the surrounding medium and from positron annihilation in
flight, and leads to a net $\sim 20$\% excess of electrons over
positrons in the shower. Askaryan's effect was revisited
with detailed simulations in the 1990's~\cite{ZHS,Alv97,Alv98} 
has recently been confirmed in accelerator measurements at 
Argonne and SLAC~\cite{Lun1,Lun2}.

The resulting coherent radio emission, the power of which rises 
quadratically with shower energy, dominates all secondary radiation 
processes (for example, optical Cherenkov) in
electromagnetic showers above about 10 PeV.
With remarkable prescience, Askaryan noted that, if there were
neutrinos above these high energies, clear natural dielectric
media such as rock salt, ice, or even the lunar regolith, might
provide the necessary large volumes needed for a suitable
neutrino target.

Spurred on by a revival of interest in techniques for radio
detection of cascades~\cite{radhep}, and by some recent efforts
in laboratory measurements of natural salt samples~\cite{chiba00},
we have made {\em in situ} measurements of
the dielectric properties or rock salt in two underground
salt excavations: the Waste Isolation
Pilot Plant (WIPP) in New Mexico, and the United Salt Corporation's Hockley
mine, within the Hockley salt dome near Houston, Texas. In the
following section we provide some background material on 
these sites and measurements of their
dielectric properties. We then report on
the two sets of experiments conducted and their results.
Section~5 presents a reanalysis of some earlier unpublished data
from ground-penetrating radar measurements in the Hockley mine.
In section~6 we discuss the implications of our results, and
our conclusions.

\section{Geological \& dielectric properties of salt formations}

\subsection{Geological properties}

Rock salt deposits are distributed throughout the world and occur 
primarily in the form of 1) bedded salts and other minerals
(known as {\em evaporites}) consisting of layers of dried solutes from 
ancient oceans; and 2) evolved salt structures which are thought
to form from the deformation of deeply bedded salt through
tectonic and buoyant forces. The latter type includes the
so-called {\em salt diapirs}: formations such as salt domes and
salt walls which involve apparent extrusion of large masses of bedded 
salt into overlying rock. This extrusion process is thought to be
driven by the fact that rock salt is generally less dense 
($\rho = 2.2$~g~cm$^{-3}$)
than most other rocks, and at the high temperatures of the 
salt beds at depths of 10 km or more, 
the salt plasticity increases~\cite{Hanna34,Halbouty}.

During the process of diapirism, through a sequence of events which is not 
well understood, the impurities in the salt tend to segregate
away from the main salt body, and the salt tends to become 
more pure than the initial bed from which it arises. In general,
where bedded salt may contain significant entrapped brine and
numerous mineral impurities, salt domes tend to produce salt
with negligible brine content and minimal impurities. The most
common residual impurity in salt domes is anhydrite (CaSO$_4$),
which occurs at the 1--5\% level in many domes.

\subsection{Dielectric properties}

\subsubsection{Pure salt}

Pure crystalline NaCl (the mineral {\em halite}) is 
known to have extremely low loss for propagation
of radio waves from frequencies of a few MHz up to 
10 GHz and more. Because of the simple ionic structure of the
NaCl crystal, there are no first-order ionic or vibrational modes that can absorb energy in this frequency range, and any absorption or scattering 
that does occur is due primarily to activation of lattice 
defects~\cite{Breck48}. Radio-frequency (RF) absorption
in dielectrics is usually described in terms of the loss tangent. 
For low-loss materials  the loss tangent is approximately the tangent of the
change in phase angle of the electric and
magnetic fields of an electromagnetic
wave with respect to a lossless medium.  It is given simply by
\begin{equation}
\tan \delta ~=~ {\epsilon'' \over \epsilon'}
\end{equation}
where $\epsilon'',~\epsilon'$ are the imaginary and real parts of the relative
dielectric permittivity.

The loss tangent is related to field attenuation coefficient $\alpha$ 
(the inverse of the distance over
which the electromagnetic field strength falls to $1/e$ of its
value) by~\cite{vonHipp}:
\begin{equation}
        \tan \delta = \left ( \left [ {2 \over \epsilon'} 
                        \left ( {\alpha c \over 2 \pi \nu } \right )^2
                        + 1 \right ]^2 - 1 \right )^{1/2}
\end{equation}
where $c$ is the speed of light and $\nu$ the radio frequency.
Conversely, the field attenuation length $L_{\alpha}$ is given by
\begin{equation}
        L_{\alpha}~=~{1 \over \alpha} = {\lambda_0 \over 2 \pi}
\left [  {2 \over \epsilon'(\sqrt{1 + \tan^2\delta}~-1)} \right ]^{1/2} 
\end{equation}
where $\lambda_0$ is the free-space wavelength of the radiation.
For the low-loss media ($\epsilon''/\epsilon' \ll 1$)
considered here, the
relationship is well approximated by
\begin{equation}
L_{\alpha} \approx 
\frac{\lambda_0}{\pi \sqrt{\epsilon'} \tan \delta}.
\end{equation}
For pure crystalline NaCl, $\epsilon' = 6.0$ and $\tan \delta \leq 10^{-4}$
over the frequency range from $\sim 1$ MHz to several GHz, and
the implied attenuation length at $\lambda_0 = 1$~m (300~MHz) is 
$L_{\alpha}\geq 1.3$~km. 

\subsubsection{Naturally occurring rock salt}

As noted above, the purity and thus the dielectric properties of natural salt deposits vary widely. Salt found in bedded evaporite deposits in North America
shows dielectric constants ranging from 5-7 and loss tangents from 
0.015--0.030 or more at 300 MHz~\cite{Annan88}, implying attenuation
lengths below 10 m, although more transparent evaporite salt
may be found occasionally in some layers.

In salt domes and other diapirs, the situation changes dramatically.
During the late 1960's and early 1970's, there was a significant
effort on the part of mining geologists to develop ground-penetrating
radar (GPR) techniques that could provide for ``look-ahead'' capabilities
in tunnel mining, to mitigate risk. Thus numerous measurements of
salt dome transparency using GPR techniques are available in the
geology literature, although the transparency results are seldom expressed
directly in terms of the loss tangent. Unterberger~\cite{Unt78}
and Stewart \& Unterberger~\cite{Stew76}
report VHF loss tangents of $10^{-4}$ for samples of 
several Gulf Coast salt dome halites, and in some cases they measured values
as low as $2 \times 10^{-5}$. Measurements of the attenuation length
{\em in situ} are not common, however. Typically GPR reports
detail the returned power of reflective regions within the salt mass,
often at great distances. In several cases, the radar systems were able
to detect reflections from the flank of a salt dome up to 1.5~km 
distant, with relatively low power radar systems (typically a few
watts peak transmitted power)~\cite{Stew76}.

Several conclusions can be gleaned from the GPR measurements of
salt domes. First, RF propagation through the salt is relatively
free from significant bulk scattering effects.
If the scattering length were short compared to the two-way propagation
distance in many of these experiments, the short pulses used in
the radar system (in some cases less than 10~ns) would not remain
coherent. This is not to say that there are no inhomogeneities
in the salt mass that can scatter radiation; these are certainly
present, but tend to be discrete and widely spaced.

Second, there appear to be no significant depolarization effects in RF
propagation through the salt. Since most transmitters used
100\% polarized radiation (either linear or circular) and received
also one polarization, the coherence of specular reflections received
in two way trips of several km could not be retained if there
were any depolarization or even polarization rotation, such as
through birefringence.

Finally, there is no evidence for any significant dispersive effects in 
low-loss salt over the frequency range from 100-1000~MHz. This
can be concluded from comparison of the dielectric constants
for many different measurements at many different frequencies in
this range; all find values very close to that of pure salt.


\section{WIPP Experiments}
The Waste Isolation Pilot Plant is primarily a repository for 
nuclear waste, but has also been directed to support 
underground experimental research. If the WIPP halite were found to have
favorable properties for RF transmission, then significant infrastructure
would be available to support development of a large neutrino detector.
The WIPP facility is operated almost exclusively on a single {\em horizon}
(or horizontal level) at 655 m depth below the local surface. The 
tunnels, known as {\em drifts}, are typically between 7--15 m wide and
about 5--6 m high. 
The average WIPP halite is 90--95\% pure, the remainder consisting
of clay, anhydrite, polyhalite, and trases of saturated brine.  
The disposal horizon is located in the 2000 feet-thick Salado
salt near Carlsbad, New Mexico.  That salt sequence, which is more
than 250~million years old, is part of a series of sedimentary
rocks filling the Permian Basin, which extends over portions of
Kansas, Texas, Oklahoma and Colorado.

Some prior measurements of the dielectric properties of four samples
of Carlsbad halite cored from a 755 m depth are given by 
Olhoeft~\cite{Olhoeft} up to a frequency of 1~MHz. 
At this frequency he found  $\epsilon' = 7.0-8.7$ and 
$\tan\delta = 0.22 -0.36$.
GPR measurements at WIPP have also been
recently done~\cite{Rogg97} at 500 MHz, however no estimates of the
attenuation length of loss tangent were reported. The goal of the
GPR was to perform high-resolution mapping of features close to
the drift walls, thus no targets deeper than $\sim 5$ m were reported.
Roggenthen~\cite{Roggpc} did report that transmission through
approximately 30 m of salt was also achieved through one or more 
columns but no estimates of the transmitted power or receiver
dynamic range were available.

We report here on radio attenuation length measurements made in 
December~2000.
Our principle measurements are derived from three four inch diameter
bore holes, separated by 22.86~m along
the ceiling centerline of room-6, Panel 2, 655~m below
the surface.   Each hole extended 6.71~m above the ceiling.
Balanced copper dipole antennas (length=27~cm, diam.=2.5~cm)
were raised into the holes using
1/2'' rigid heliax cable
at a nominal height of 5.49 m above the ceiling corresponding to
the center of the purest accessible halite layer.
Note that the ceiling and floor of the drifts follow the geological
strata so that the antennas at the same height above the
ceiling are in the same layer.  The signal pulse used for these
and the following measurements in the Hockley mine consisted of
a 10-100 ns pulse, modulated with frequencies from 90-500 MHz.

The signal was 
sent to a dipole in one of the boreholes which transmitted into
the halite.
The signal was received by a dipole inserted into another of the 
boreholes, then amplified by two cascaded, but physically separated, 
broadband amplifiers for a total gain
of $\sim57$~dB.  The output of the second amplifier was filtered
by an appropriate bandpass filter about the center frequency
to improve the signal/noise ratio.
The received signal was recorded by a
digital oscilloscope.   When the transmitting
dipole was moved from the end to center
hole, the acquisition system (amplifiers, cables, filters etc.)
were kept identical so that the overall system gain did not need to
be known.

The field attenuation factor, $\alpha$ is determined from
the received voltages, $V_1$ and $V_2$ at two distances,
$d_1$ and $d_2$ by,
\begin{equation}
\frac{V_2}{V_1}\frac{d_2}{d_1} = \exp (-\alpha(d_2-d_1)).
\end{equation}

\begin{figure}[t]
\begin{center}
\leavevmode
\epsfxsize=4.0in
\epsfbox{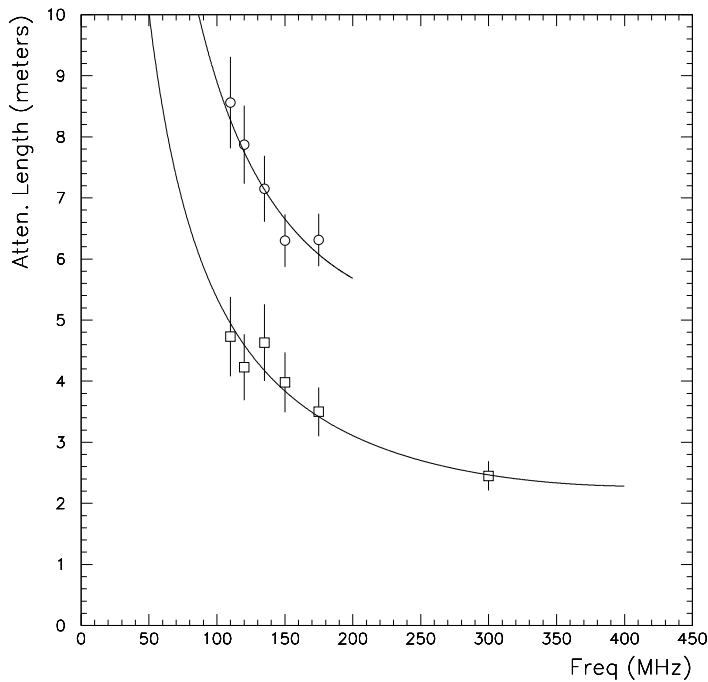}
\caption{Measured field attenuation length versus frequency from
the two sets of three holes.}
\label{attenvsfreq}
\end{center}
\end{figure}

Attenuation lengths vs. frequency at this stratum are summarized
in fig.~\ref{attenvsfreq}.
Only signals between
110~and 175~MHz coupled well enough to be easily observed at the far hole.
At 300~MHz, the pulse was extracted from the noise using
a cross-correlation with the transmitted pulse shape.
The assigned uncertainties are due to a 15\% r.m.s. 
azimuthal asymmetry of the dipoles and 0.5~m distance uncertainty.
The data are fairly well described by a loss-tangent that is independent
of frequency.   We do, however,  
see some evidence for a slow decrease in loss tangent
versus frequency as would be consistent with the presence of
some brine at these frequencies~\cite{vonHipp} so we fit the data to
a frequency-dependent loss-tangent:
\begin{equation}
\tan \delta (\nu) =  a + b\frac{\nu -\nu_0}{\nu_0},
\end{equation}
where $a$ would be the loss-tangent if it were frequency independent and
$b$ is a small correction.   We chose $\nu_0=200$~MHz in the
parameterization to minimize the correlation between $a$ and $b$.  
The lower set of data in fig.~\ref{attenvsfreq}, 
which includes a 300~MHz point, 
fits to:  $a=0.0627\pm 0.0031$ and
$b=-0.020\pm 0.008$
In the upper set of data, which did not include a 300~MHz point,
$a$ and $b$ were highly correlated but we find 
$a=0.0343\pm 0.0036$ and $b=-0.019\pm 0.012$.
Both sets are thus consistent with perhaps a slow decrease of loss tangent
at increasing frequency, as expected from the behavior 
of the loss tangent of brine.

We anticipated that the attenuation length would depend on geological 
layer since
the relative amounts of impurities, clay and anhydrite, 
vary
significantly. 
Measurements vs. height
were made at 150~MHz, where the antenna coupling was observed to
be the most efficient.  The attenuation
length measurements were made between only two boreholes and
are normalized to the value measured at 5.5~m.
The results are summarized in Table~\ref{attenvsheight}.  Despite the
changes in attenuation length, 
less than 0.5\% variation in the index of refraction for the
various layers was observed.

The delay of the 150~MHz pulse train 
over 23~m vs. 45~m gave the index of refraction
to be $2.82\pm0.03$.  This may be compared with the measured
value for pure salt, 2.45.  
Since both anhydrite and clay have lower indices of refraction than
halite, the
difference may be due to moisture known to be trapped in
the clay-rich layers of the salt.
Water has an index of 
refraction of about 9 at  these frequencies and a very high
loss tangent, so its presence may be the cause of
both the short attenuation length and the anomalously high
index of refraction.

\begin{table}
\begin{center}
\caption{Attenuation length versus
stratum as measured in height of antenna
above ceiling.  These measurements are normalized to the
value measured at 5.5~m.  Uncertainties are systematic as described
in the text.}
\vspace{0.5cm}
\begin{tabular}{c|c|c}
height above ceiling&impurity present&attenuation length\\
(meters) & ~  & (meters)\\
\hline
3.7 & clay & $3.4\pm 0.5$\\
4.6 & anhydrite stringers& $4.1\pm 0.6$\\
5.0 & anhydrite stringers & $6.3\pm 0.9$\\
5.5 & least anhydrite& $\equiv 6.3$\\
5.9 & some anhydrite & $3.6\pm 0.5$\\
6.4 & some anhydrite & $4.5\pm 0.7$\\
6.7 & much clay/some anhydrite & $4.2\pm 0.6$\\
\end{tabular}
\label{attenvsheight}
\end{center}
\end{table}

\section{Hockley mine experiments}

The Hockley salt dome was discovered in 1905 due to gas seeps and
other evidence for trapped oil reservoirs in the area. Some
oil production has continued up to recent times but the dome
is not highly productive for oil. In 1930 the Houston Salt Company
drilled a shaft in the northeast part of the dome and began
mining salt at a depth of 460~m in 1934, but ceased operation within
several years. The United Salt Company acquired the mine in 
1946 and has continued to mine salt until the present\cite{Hluch73}. 
The current mine covers an area of several square km in a grid of
10-15~m wide by 5-8~m high drifts,
and rectangular columns typically 30~m by 40~m in cross section.

The top of the salt structure begins at a depth of about 300~m
below the local surface. The dome is roughly elliptical in 
horizontal cross section,
with major and minor diameters of 3.6~km by 2.9~km at the mine level.
The cross section of the dome continues to grow with depth to at
least 2~km, where it is 4.4~km by 3.3~km.
The salt structure itself is thought to
extend down to 10~km depth, with an inverted teardrop shape that narrows
at the base. The estimated volume of salt in the dome is
of order 80 km$^3$~\cite{Ritz36}, with of order 20-25 km$^{3}$ 
contained within the top 3~km of the dome.~\cite{Halbouty}

We made measurements in the Hockley mine in June 2001.
For the Hockley mine experiments, boreholes were not available,
and measurements were made primarily with two antennas: a half-wave
dipole that was tuned to peak at 150 MHz in salt, and which worked
also at the full-wave resonance at 300 MHz. This antenna was found to
behave reasonably well if it were in contact with the salt surface,
although we could not measure the modified beam pattern of the antenna.
We also made use of a UHF 4-bay bowtie antenna with a ground-plane 
and approximately 12 dBi of gain for measurements where the antenna was
aimed into the salt from an external position. This latter antenna
was used primarily at a frequency of 750 MHz.

In all cases, our goal was to was to make measurements over several
different distances with the same antenna and cable configuration,
so that we could make use of relative measurements which would not
require absolute antenna calibration. Our ability to do this was 
constrained by the geometry of the region of the mine where we
were able to operate, and we found in several cases that 
the local shape of the wall and other anomalies 
prevented our using some of the positions.

\begin{figure}
\begin{center}
\leavevmode
\epsfxsize=4.0in
\epsfbox{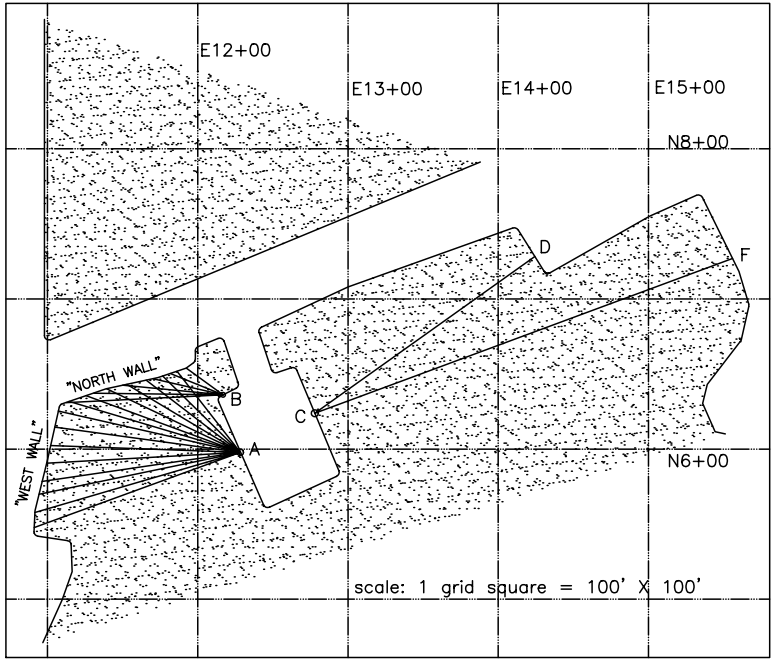}
\vspace{0.5cm}
\caption{Layout of the region of the mine where our 
Hockley measurements were made. Typical rays for the measurements are
shown.}
\label{hockmap}
\end{center}
\end{figure}

Figure \ref{hockmap} shows a layout of the region of the
mine where we made our measurements. The grid is in mine coordinates
which are $100'$ squares aligned along compass directions.
The primary positions of the receive antenna are indicated by
the letters A,B,C. Most of the measurements were made with the
transmitting antenna along either the north or west walls as
indicated. However, we also made some tests using the longer
distance to the east, between positions C \& D (about 40~m)
and C \& F (about 90~m). In these latter cases the mine drift 
was at a 4--6~m lower elevation than the receive antenna, and
some difficulties were encountered with questionable
emplacement of the antenna,
due to the presence of fractures
in the salt wall of the drift. Thus although we were able to 
successfully transmit through the salt at both 150~MHZ and 300~MHz 
at these locations, the data were not repeatable enough to give
reliable results.

One of the recurring
difficulties was due to the fact that while obtaining a range
of distances for the measurements, we could not retain
normal-incidence angles for the transmission or reception
antennas. For vertical dipole measurements, the beam pattern has
nominal azimuthal symmetry. However, the presence of the salt-air
interface modifies this pattern significantly, particularly
for rays that are far from normal incidence.  

\subsection{Attenuation measurements}

\subsubsection{Absolute field strength measurements.}

As can be seen from Fig.~\ref{hockmap}, a large number of
measurements were made along both the north and west walls, with distances
ranging from about 12 to 45 m. To obtain an indication of whether there
were any overall gross systematics in the transmission or receiving of
the signals, we here perform an analytical estimate of the expected
received signal compared to the pulse which was measured on
transmission. Although we found large scatter in the measured data,
this procedure will provide a first-order evaluation of the
attenuation in the salt and will indicate whether there is a subset of higher
quality measurements that can be reasonably used to improve this estimate.

The Friis formula for the relationship between 
transmitted ($T_x$) and received ($R_x$) power in an antenna is given by 
\begin{equation}
{{P_{R_x}}\over{P_{T_x}}} = {{{A_{T_x}}{A_{R_x}}}\over{{\lambda^2}{R^2}}},
\end{equation}
where $A$ is the effective area of the transmitting or receiving antenna, 
$\lambda$ is the wavelength in salt
and $R$ is the distance between the two antennas.

Recasting this equation in terms of the voltage measured:
\begin{equation}
{{V_{R_x}}\over{V_{T_x}}} = \sqrt{{{A_{T_x}}{A_{R_x}}}\over{{\lambda^2}{R^2}}}.
\end{equation}
Noting that $A_{T_x} = A_{R_x}$, we have
\begin{equation}
{{V_{R_x}}\over{V_{T_x}}}R = {A\over\lambda}.
\end{equation}

The effective areas of a half-wave dipole and full-wave dipoles are 
$0.13 \lambda^2$  and  $0.048 \lambda^2$, 
respectively.~\cite{Kraus88}
The dipoles used at 
Hockley are half-wave at 150 MHz and full wave at 300 MHz, so at 150 MHz,

\begin{equation}
{{V_{R_x}}\over{V_{T_x}}}R = 0.13{\lambda},
\end{equation}
while at 300 MHz,
\begin{equation}
{{V_{R_x}}\over{V_{T_x}}}R = 0.048{\lambda}.
\end{equation}

\begin{figure}
\begin{center}
\leavevmode
\epsfxsize=4in
\epsfbox{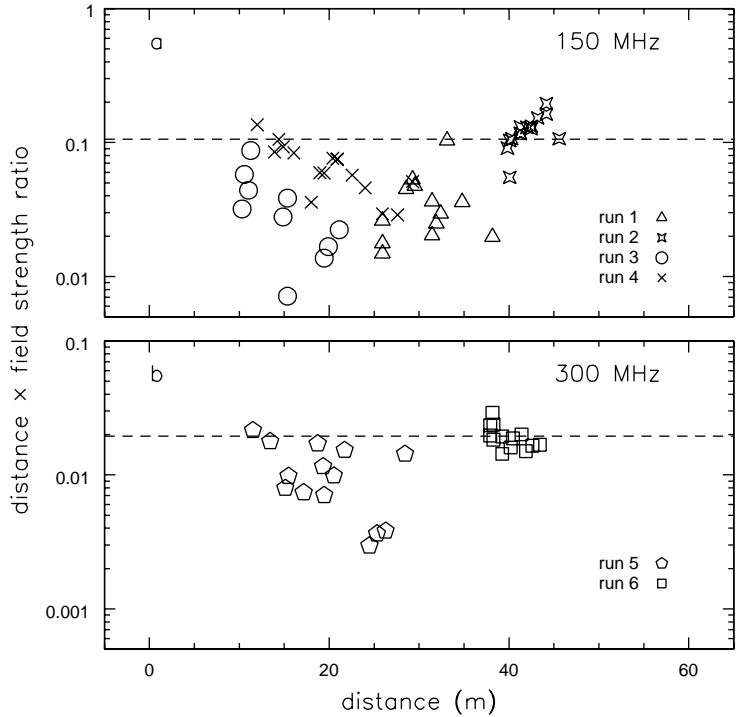}
\caption{Field strength measurements for all of the data taken 
along the N and W walls at (a) 150 \& (b) 300 MHz.  The dashed line is the 
expected value without attenuation.}
\label{scatter}
\end{center}
\end{figure}

Figure~\ref{scatter} shows the quantity ${{V_{R_x}}\over{V_{T_x}}}R$ 
plotted for all of the data
at 150 and 300~ MHz.  The dashed line shows the 
absolute value for ${{V_{R_x}}\over{V_{T_x}}}R$ given by the Friis formula.
The data were taken in six distinct ``runs'' with one of the dipoles
kept in a fixed position and the other dipole was moved.  The 
runs with the least repeatability were taken with the fixed dipole
near a large irregularity in the surface of the wall.
The data points are corrected for cable attenuation, amplifier gain,
and for reflection losses. At 150 MHz, $\sim$30\% of the power sent to the
transmitting antenna is reflected back.  At 300~MHz, the reflected
power was not directly measured, so the same reflected power is
assumed for 300~MHz.  Note that the Friis formula assumes a dipole
pattern, which is not completely accurate in our case since the
antenna pattern was half in salt and half in air. We have made no
correction for this effect.

The data points are consistent, within an order of magnitude, 
with the value given by the Friis formula, which assumes no attenuation 
due to 
the medium.  Note that the points at the farthest distance were taken with
the best geometry and showed the least scatter. By virtue of having the
greatest separation, these data also have the best
sensitivity to attenuation losses. These samples show no measurable
attenuation.
Therefore the data are consistent with a very long attenuation length, 
well in excess of 40 meters.

\subsubsection{Relative field strength measurements.}

To improve the precision of our estimate of the attenuation length,
we have excluded runs 1 and 3 of the 150~MHz data due to questions 
about coupling systematics, and we have excluded the top and bottom
10\% outlying samples of the remaining data.
We have grouped the data in
appropriate range bins and averaged the received pulse power for
each range bin. No correction has been made for non-normal incidence 
angles. After averaging we then convert back to the mean field strength
by taking the square root of the result. The fractional power variance
of the averaged data provides an estimate of the precision of the field 
measurement.

As noted above, in the far field
the ratio of the received ($R_x$) to transmitted ($T_x$) 
voltage in an antenna should behave as
\begin{equation}
{V_{Rx} \over V_{Tx}} = {\kappa \over d}\exp(-\alpha d)
\end{equation}
where $d$ is the distance between the transmitting 
and receiving antennas, $\alpha$ is the field attenuation coefficient, 
and $\kappa$ is an unknown
constant scale factor which accounts for the system losses.
In Fig.~\ref{atten}(a) and (b) we plot  $(V_{Rx}/ V_{Tx}) \times d$ 
on a logarithmic scale as
a function of distance for 150 and 300~MHz, 
where we have now normalized to the mean power. The plot also shows the 
results of fits to the exponential attenuation factor (here
expressed in terms of its inverse, the attenuation length). Note
that the error bars from the exponential fit
are markedly asymmetric and that values below
zero for either attenuation coefficient or attenuation length are
unphysical.  The uncertainties are given as the range of the 67\% confidence
interval. Thus, for Fig.~\ref{atten}(a), the 150~MHz attenuation
length is consistent with 34~m at the low edge of the confidence
interval, but is inconsistent with values below 20~m at the $\geq 95$\%
confidence level.

\subsubsection{Attenuation ratio for 300 to 150 MHz.}

Because the 300 MHz and 150 MHz measurements were made with the
same antenna used in full- and half-wave dipole mode, a
relatively simple relationship holds between the attenuation
measurements for the two frequencies.
If the loss tangent is nearly constant with frequency, which we expect from 
our previous measurements at WIPP, then the field attenuation 
coefficient  is proportional to the frequency. 
That is, the attenuation length at 300 MHz is half of the 
attenuation length at 150 MHz. Then

\begin{equation}
{V_{Rx^{300}}\over V_{Tx^{300}}}
\left [{V_{Rx^{150}}\over V_{Tx^{150}}} \right ]^{-1} = \exp(-\alpha_{150}~d)~.
\end{equation}

\begin{figure}
\begin{center}
\leavevmode
\epsfxsize=5in
\epsfbox{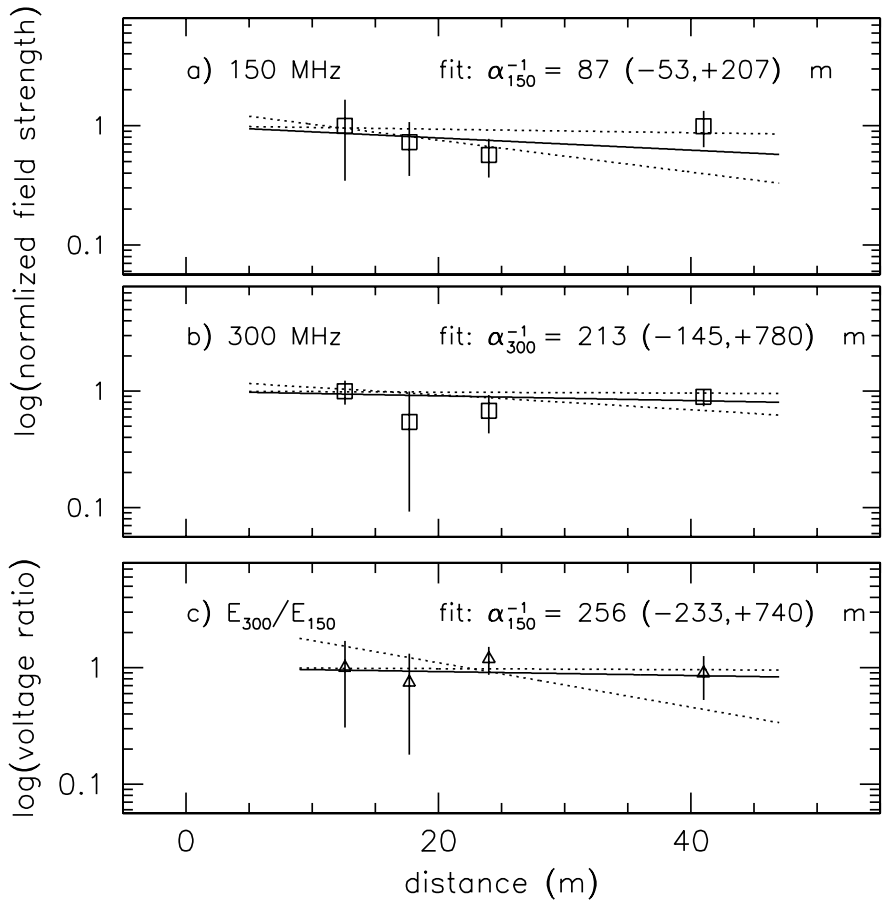}
\caption{Plots of binned field attenuation measurements. (a) The data at
150 MHz, with exponential fit shown as a solid line, including 67\% confidence
interval in dotted lines. (b) Same as (a) for 300~MHz. (c) Ratio of
(b) over (a), which gives a partially independent estimate of the
150~MHz attenuation length, shown as a fitted line as in (a) and (b).}
\label{atten}
\end{center}
\end{figure}

Thus by taking the ratio of the two relative attenuation measurements, we
obtain a partially independent measure of the attenuation at 150 MHz,
under the assumption of a constant loss tangent.
Note that we found a slight deviation from a constant loss tangent at
WIPP.  We attribute that deviation to the brine and other
impurities trapped in the salt
at WIPP that is not present at Hockley. Measurements of Texas salt
dome halite in the early 1970's~\cite{Unterberger70} do report a
slow decrease of the loss tangent with frequency for low-loss salt.
If we chose to apply
a correction for this effect to these data, it
would have a small effect ($\sim$20\%) on the reported attenuation length
derived from this ratio.
Figure~\ref{atten}c shows this
ratio on a logarithmic scale as a function of distance. For these
data, the slope of the logarithm is a measure of the 
attenuation length as described
above.  A fit of the data to a falling exponential 
indicates a large value for the 
attenuation length; however the fitted uncertainties are also large,
and the result is consistent with the previous results. 
It is evident that a 150 MHz attenuation length much greater
than 40 m is favored by the data.

\subsubsection{Attenuation at 750 MHz.}

We also made measurements of transmission at
750 MHz using a commercial UHF antenna consisting of four phased bowtie
antennas.  Because this antenna array
had a much narrower beam than the dipoles we used, we were affected 
more by the difficulty of alignment of the antenna beam for
the transmit and receive antennas.   Because of wide variations in
the received power, we chose in this case to only
use only the highest $\sim 50$\%
of the measured values to reduce the otherwise significant scatter.
(This was only done for the 750~MHz data which were taken with a high
gain antenna.)
This approach is in general justified since beam misalignments and 
coupling problems produce only losses and lensing effects can be
neglected here~\cite{Hols72}).  The data were then power averaged within
appropriate range bins. In this case, we have combined data
from two separate runs which are offset by an unknown
normalization, by rescaling each data set to its mean
power.  

\begin{figure}
\begin{center}
\leavevmode
\epsfxsize=4in
\epsfbox{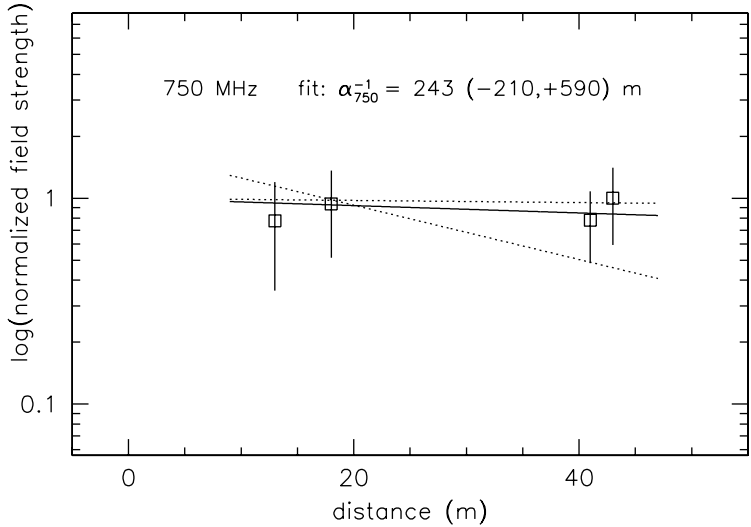}
\caption{Similar to previous figure, for 750 MHz data.}
\label{att750}
\end{center}
\end{figure}

The results of the 750 MHz analysis are shown in Fig.~\ref{att750}.
Again there is little or no evidence for attenuation, and the
fitted value shown is not strongly constrained by the data.
It is remarkable that despite the factor of 2.5 increase
in frequency from 300~MHz, the
data are still consistent with an extremely long attenuation length.

\subsection{Polarization measurements.}

A half-wave dipole, which produces linearly polarized
radiation with electric field vector aligned with the dipole axis,
typically provides about 20 dB of cross-polarization rejection.
We made measurements at 150~MHz of the cross-polarization leakage 
through a distance of 45~m of the salt. The net
cross-polarized power observed was $(1.55 \pm 0.1)$\% of the
co-polarized power. This is
consistent with no polarization leakage since this is
within the cross-polarized rejection limits of the antennas.

We note that this measurement also indicates a lack of significant
birefringence of the salt, since any rotation of the plane
of polarization would appear as a noticeable cross-polarization
leakage. Using the same cross-polarized power above, the implied
limit on the rotation angle $\Psi$ of the plane of polarization is 
$\Delta\Psi \leq (0.16 \pm 0.01)^{\circ}$~m$^{-1}$. Thus the
phase difference between two circularly polarized modes will
not exceed 1 radian in 360~m of propagation distance in the salt.

\subsection{Noise environment}

The overburden of rock above the salt dome should insulate the
environment well from terrestrial radio-frequency interference (RFI).   
We measured the noise environment at the Hockley site
using the same short dipole used at 150~MHz.  
We calculated that the system temperature of our apparatus
was 770~K including front-end 
amplifier noise (263~K) , cables (1.2~dB at 310~K), and 310~K 
salt filling the antenna aperture.
We observed no departure from uniform power in the power spectrum.   
We could not observe any difference between observing the
salt through an antenna versus a 50~$\Omega$ load at the front-end
amplifier input.
We could have detected excess power from the salt (above its
blackbody temperature) of order 100~K if it were present.
Hence we conclude the noise environment may be 
characterized as fluctuations of a 310~K blackbody spectrum.  On 
rare occasions
we could see clear RFI due to the use of walkie-talkies by local
mining crews.   Such events in a salt detector
would be easily removed by offline analysis but could cause a high 
trigger rate.  To keep the trigger rate low, a salt detector in the mine 
might require tuned notch filters for the communication
frequencies in use at the facility.

\section{Analysis  of  unpublished GPR data from Hockley}

As we have noted earlier, there is a significant body of
geophysics literature on ground-penetrating radar measurements
of salt formations. In particular, J. Hluchanek, under the supervision
of R. Unterberger, completed a master's thesis using
GPR measurements in the Hockley mine~\cite{Hluch73}. In this
section we analyze some of these results with the goal of 
obtaining an independent estimate of the attenuation length.

The Texas A\&M GPR system used a beamed Yagi-Uda antenna at 440~MHz.
Since the primary goal of these measurements was to establish the
utility of GPR for forward-looking assessment of the path of a
mined tunnel, almost all of the data consist of measurements of the
amplitudes of reflections from inhomogeneities within the salt,
at several positions within the mine, and at various distances
within the salt mass, ranging from about 50~m to over 350~m from
the transmitter. Having measured the salt index of refraction 
($2.45\pm0.05$, consistent with pure salt),
Hluchanek was able to establish the target distances to a precision
of several m, and estimate the received power over a dynamic range of 
about 50~dB. No absolute measurements of the transmitted or 
received power were given, except that the transmitter itself
operated at about 3~W peak power.

\begin{figure}
\begin{center}
\leavevmode
\epsfxsize=4in
\epsfbox{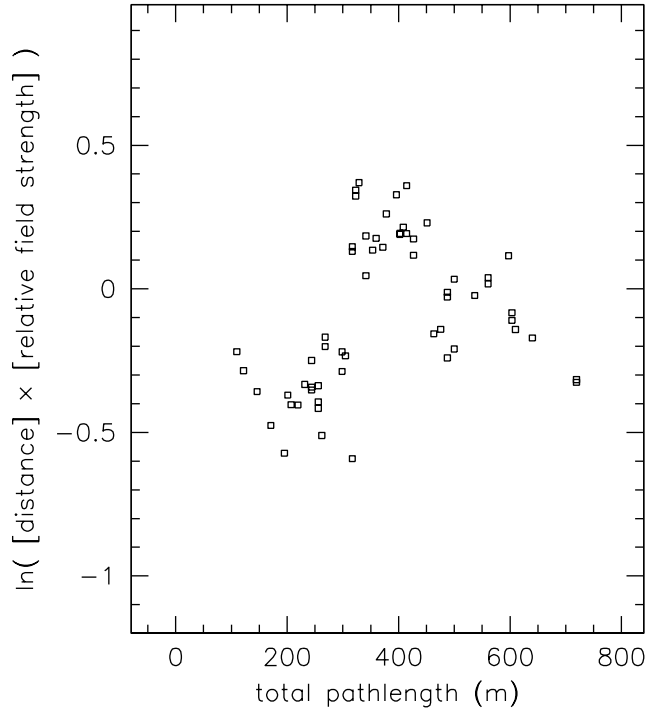}
\caption{Measurements of relative received field strength from 
radar targets in the Hockley mine.
The data have been boxcar smoothed with over $\pm 5$ adjacent data points
to reduce the variance.}
\label{hockgpr}
\end{center}
\end{figure}

Given these data, we arrive at a first order estimate of the
attenuation length by assuming that the ensemble
of reflective inhomogeneities has a well-defined mean reflectivity
with a quasi-normal distribution around this mean. Since,
when a target was seen in the radar return, the antenna was then
adjusted to maximize the return power, we can in general ignore
the antenna beam effects. Hluchanek also made Snell's
law corrections for cases of non-normal incidence in the geometry.

Figure~\ref{hockgpr} shows a plot of the field strength data,
multiplied by the two-way propagation 
distance to remove the radial loss factor. Here
we have smoothed the power measurements with a 5-sample
boxcar average to reduce the scatter, 
and then converted to field strength by
taking the square root. 
It is evident that the data do not follow
a simple exponential as would be the case for a uniform
reflectivity of the targets. A simple fit to all of the data in this case
would yield no measurable attenuation. However we can calculate
an attenuation length by
fitting only targets at distances above 300 m, treating them as
a uniform ensemble.  This approach
is somewhat more conservative than fitting the entire data set,
since the data below 300 m would tend to force a much longer
attenuation length.

\begin{figure}
\begin{center}
\leavevmode
\epsfxsize=4in
\epsfbox{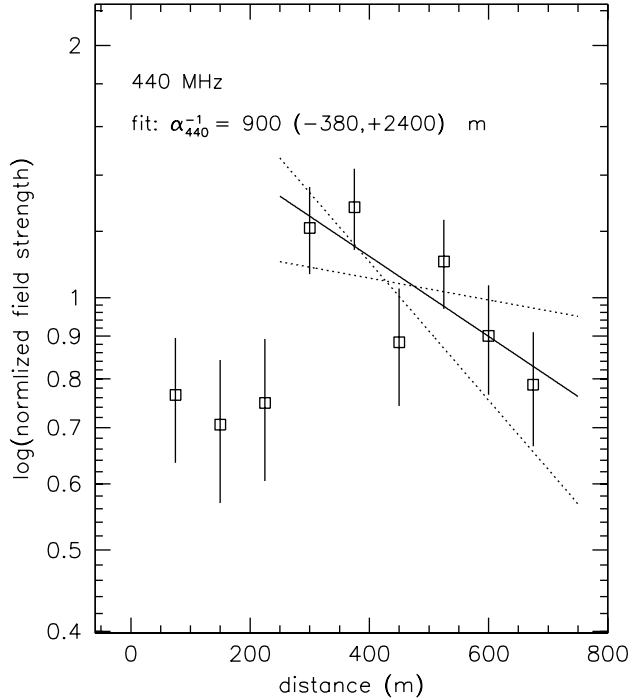}
\caption{A plot of the GPR data, now binned in 75 m bins.
The fitted lines are the fit of only the data above 300 m, and
the 67\% confidence interval fitted lines are also
shown. Quoted uncertainties are statistical only, and do not include
potentially significant systematic errors.}
\label{hockgpr2}
\end{center}
\end{figure}

Fig.~\ref{hockgpr2} shows the results of this approach. Now
we have binned the data in 75 m bins, with uncertainties assigned
according to an estimate of the power-weighted variance of
the data in each bin. The resulting fit is plotted as in the
previous section. The uncertainties shown are only statistical;
clearly there are several possible systematic uncertainties that could
impact this analysis. However, the indications are again that
the attenuation length in the VHF to UHF regime in the Hockley
dome is of order several hundred meters at least. Such
long attenuation lengths are in fact quite difficult to
accurately measure under any conditions, whether in the
laboratory or {\em in situ}.

\section{Discussion}

We have established that naturally occurring salt formations
exist with extremely long radio attenuation lengths, comparable or
better than the clarity of pure water or ice at optical wavelengths.
We should note here that we have followed the RF convention~\cite{vonHipp}
and expressed attenuation lengths in terms of the field attenuation
rather than power or intensity attenuation, as is more common
at optical wavelengths.  However, using the separate definitions
is appropriate for comparing techniques, since at radio frequencies the
detection is coherent and signal strength increases linearly
with field strength. In optical detection of Cherenkov radiation,
the fields sum incoherently, and the signal strength increases
linearly with intensity rather than field strength.

A natural dielectric material that has RF attenuation lengths
comparable to that of salt is clear glacial or polar shelf ice.
As we have noted in the introduction, the RICE experiment has already
begun to exploit this property of ice in the search for 
neutrino interactions~\cite{fri96,Bes99,Bes01}. 
It is thus of interest to
compare and contrast the properties of ice and salt for this
purpose.

\begin{figure}
\begin{center}
\leavevmode
\epsfxsize=5in
\epsfbox{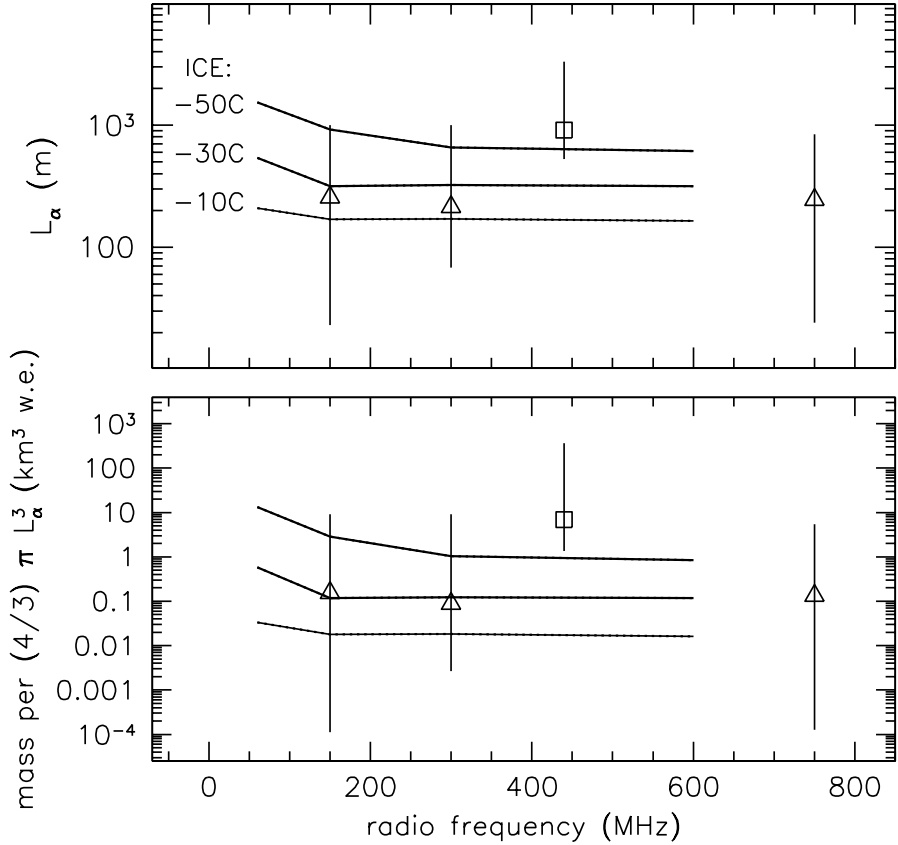}
\caption{Top: Attenuation lengths of antarctic ice for various
temperatures, and for the salt results reported here. Bottom,
enclosed mass (in water-equivalent cubic km)
within one attenuation length radius for ice and salt.   Triangles
are the data taken by our group; squares are the re-analyzed GPR
data.}
\label{saltice}
\end{center}
\end{figure}

In Fig.~\ref{saltice}(a) we show curves for measured attenuation
lengths in ice~\cite{iceatten} and the results reported here,
using the longer estimated attenuation length at 150 MHz,
from the ratio data.
The attenuation length in ice is strongly temperature dependent.
The temperature profiles in antarctic shelf ice and in glacial
ice can vary significantly from site to site and often show
an inversion at increasing depths.  However, one can typically
assume that a range of order 1~km of depth of ice with temperatures of 
$-20^{\circ}$ to $-30^{\circ}$C or colder can be found at most sites. 

If we ignore for the moment the practical limits to
contiguous volume of material, Fig.~\ref{saltice}(b) shows the
contained mass per volume with a radius of one attenuation length
for ice and for the salt measurements we report here. It
is evident that the significantly greater density of salt
compared to ice (a factor of 2.4) leads to quite similar 
detection masses for salt and ice at $-30^{\circ}$C. It is notable that
in all cases the mass per attenuation volume exceeds 0.1~cubic
km water equivalent for salt, and for ice below $-30^{\circ}$C.

We note that the higher density of salt compared to ice, as well as
the higher index of refraction, leads to several other differences
in the behavior of Cherenkov emission from a cascade within the
medium. Cherenkov power $W_c$ depends on the index of refraction $n$ as
\begin{equation}
W_c \propto 1 - {1 \over n^2 \beta^2}
\end{equation}
where $\beta$ is the particle velocity with respect to light speed.
For ice at RF wavelengths, $n\simeq 1.6$, whereas for salt $n=2.45$.
Thus the Cherenkov output power is increased by 36\% for $\beta=1$,
and there is an additional increase due to the fact that the 
Cherenkov threshold is $\sim 50$\% lower than in ice, so 
particles of lower energy can contribute to the RF emission.

Because of the higher density, the cascade is also more compact by
roughly the inverse of the density. Thus there is a decrease in 
the total tracklength of the shower and thus the Cherenkov power
also decreases by roughly a factor of 2.4. For the same reason,
however, the RF coherence of the shower is maintained to higher
frequencies by roughly the same ratio and thus the coherent power
output, which scales as $\nu^2\Delta\nu$~\cite{ZHS} will substantially 
offset the loss due to shorter tracklengths.   Concerns over Fresnel
effects would also be correspondingly reduced.~\cite{ralston}

A final note on RF propagation in ice concerns possible
birefringence. Because of the complex
crystal structure of ice, it is not surprising that both
depolarization and rotation of the plane of polarization 
have been observed in both glacial 
and Antarctic ice~\cite{iceatten}. 
At 440~MHz, the fractional difference in the index of refraction
for the ordinary and extraordinary modes was found in some
cases to exceed $5 \times10^{-4}$, implying a rotation of
the plane of polarization of $\Delta\Psi\simeq 0.45^{\circ}$~m$^{-1}$,
about 3 times higher than the limit we measured for Hockley salt.
(Recent gain calibration of the RICE detector may indicate less 
birefringence, however.~\cite{riceicrc})  
Although measurements of polarization properties in salt are 
less complete, the lack of any evidence for strong birefringence,
combined with the simple cubic lattice structure of the
basic NaCl crystal leads to an expectation that salt may
be superior to ice in this respect. 
Low birefringence could have
importance for a cascade detection system since there is
significant potential gain in event reconstruction
if the polarization of the radio emission from
the cascade can be made.~\cite{Lun2}

\section{Conclusions}

We have made initial measurements of the radio frequency attenuation
lengths in natural halite in two underground salt excavations,
the WIPP facility in
New Mexico and the Hockley mine in Hockley salt dome in south Texas.
We find that WIPP halite is quite lossy, due most likely to entrapped
brine and other impurities, 
and is not suitable for a large scale detector for high
energy neutrinos.

The salt in the Hockley mine is, by contrast, extremely transparent
over the range from 150 to 750 MHz, with probable VHF and UHF
attenuation lengths of
several hundred meters or more.  These long attenuation lengths
are supported by several separate analyses: absolute 
field intensity, electric field measurements vs. distance, relative 
measurements vs. distance, and GPR data.  The obtained 
values are especially
long when one considers the density of salt relative to water.
In addition, there is no apparent depolarization or
significant scattering of the signals over a 45 m distance.  The noise
environment appears to be extremely quiet, consistent with a 310K black-body.
Thus salt appears to provide a suitable medium for potential neutrino
detectors using embedded antennas in a manner similar to that of
the RICE experiment in ice at the south pole.

To illustrate the potential power of a saltbed neutrino detector, we
assume a $10\times 10\times 10$ antenna 
array on a 200~m grid spacing, with center
frequency at 150 MHz and a 50\% bandwidth, and a 300~m attenuation length
at 300~MHz with a constant loss tangent with frequency as expected for
salt. Such an array has an instrumented volume of about 8~km$^3$ and would
easily fit within the top 3~km of a salt formation such as the Hockley
dome. A Monte Carlo simulation of this array indicates that we would
detect (4 antennas hit with a voltage SNR of 4~$\sigma$)
of order 10 events per year from the minimal GZK neutrino 
flux~\cite{sse01}
and up to 50~events per year if the flux is the
maximum allowed value. For other models of high energy neutrino fluxes,
we expect of order 5-10~events per year from gamma-ray bursts~\cite{wbgrb}
depending on source evolution,
and over 400~events per year from a representative AGN neutrino 
model~\cite{mannheim}.
The rate of such thermal coincidences over the course of a year is
a few tens of events which would be easily removed by offline event
reconstruction.   The overburden of radio-absorbing
rock protects the array from man-made radio-frequency interference.

Further work remains to be done to make more precise measurements of
these remarkably long attenuation lengths, but the salt appears
to present a viable medium for large calorimetric detectors in the
mass range of 10-100 cubic kilometers of water-equivalent mass to detect
cosmic neutrinos.

\section*{Acknowledgments}

We thank Roger Nelson, Norbert Rempe, 
Rey Carrasco, and other members of the WIPP staff for
invaluable assistance with operations in the mine.
We thank Dennis Bradley, Michael Nigh, and Alan Simon for
their generous help in our Hockley mine measurements,
and United Salt Corporation for their support of this
endeavor.
This work has been 
supported in part at UCLA by the U.S. Dept. of Energy, in particular its
Advanced Detector Research Program, and by the National Science
Foundation.
This work has been performed in part at the Jet Propulsion
Laboratory, California Institute of Technology, under contract with
the National Aeronautics and Space Administration.
The Stanford Linear Accelerator Center is supported by
the U.S. Department of Energy, with work performed under
contract DE-AC03-76SF00515.

\end{document}